\documentclass[usenatbib,usegraphicx]{mn2e}

\newcommand{\ltabout}{\la}
\newcommand{\sbar}{$\bar{S}$}
\newcommand{\dbar}{$\bar{D}$}
\newcommand{\nmove}{$n_\mathit{move}$}

\title[Automated classification of lightcurves]{The automated classification of astronomical lightcurves using Kohonen self-organising maps}

\author[D.~R.~Brett, R.~G.~West \& P.~J.~Wheatley]{David R. Brett,
Richard G. West\thanks{Contact: rgw@astro.le.ac.uk} and Peter J. Wheatley\\
Department of Physics \& Astronomy, University of Leicester, 
University Road, Leicester LE1 7RH}

\begin{document}

\maketitle

\begin{abstract}
We apply the technique of self-organising maps \citep{Kohonen90} 
to the automated classification of singly periodic astronomical lightcurves. 
We find that our maps readily distinguish between lightcurve types in both 
synthetic and 
real datasets, and that the resulting maps do not depend sensitively on the 
chosen learning parameters. Automated data 
analysis techniques are likely to be become increasingly important as the 
size of astronomical datasets continues to increase, particularly with 
the advent of ultra-wide-field survey telescopes such as WASP, RAPTOR and ASAS.
\end{abstract}

\begin{keywords} 
methods: data analysis -- 
techniques: miscellaneous -- 
astronomical databases: miscellaneous -- 
surveys -- 
binaries: general -- 
stars: general.
\end{keywords}

\section{Introduction}
Ultra-wide-field optical surveys are now capable of routinely
acquiring high-precision time-resolved photometry of large samples of 
celestial objects. Such projects include 
the All Sky Automated Survey, ASAS \citep{Pojmanski97,Pojmanski02};
the Robotic Optical Transient Search Experiment, 
ROTSE \citep{Akerlof00,Wozniak04}, 
the Rapid Telescopes for Optical Response, RAPTOR \citep{Vestrand02},
Stellar Astrophysics and Research on Exoplanets, STARE \citep{Alonso03}
and
the Wide Angle Search for Planets, WASP \citep{Kane03,Street03}.
The largest of these projects are already capable of monitoring around 
$10^8$ objects. To be of practical use it is
important that the lightcurves acquired by these projects can be 
classified automatically, so that a human user is able to filter the 
available data to discriminate lightcurves of interest. 

Previous attempts at automated classification have focused on 
singly-periodic variable stars and have tended to rely on 
an analysis of parameters derived from the lightcurves, such as period 
and Fourier co-efficients. 
Classification has been achieved by identifying specific
regions in the multi-dimensional space defined by these parameters that 
correspond to different classes of object. Examples of this approach include 
analysis of ROTSE data by \citet{Akerlof00} and analysis of ASAS data 
by \citet{Eyer02} and \citet{Pojmanski02}. 

In this paper we present an algorithm designed to classify 
singly-periodic variable stars based directly 
on their unparameterised folded lightcurves.  
Our algorithm uses an artificial neural network, and is capable of forming
clusters of lightcurves with similar shapes within an arbitrary
$N$-dimensional space. 
In Sect.\,\ref{sec-algor} we describe our algorithm in general terms, and in 
Sect.\,\ref{sec-perform} we describe and discuss its application to real 
and synthetic 
test data sets. Section\,\ref{sec-perform} includes a discussion of the 
choices made in a specific implementation of the algorithm.
Finally, in Sect.\,\ref{UmatSect}, we describe and test methods designed to 
identify clusters of lightcurves in our output maps. 

\section{The algorithm}
\label{sec-algor}

The algorithm we have developed utilises a form of neural network,
namely the self-organising map scheme due to \citet{Kohonen90}. We are
motivated in this choice by two desirable properties of Kohonen maps:
simplicity of implementation and their ability to learn in an
unsupervised r\'{e}gime. Unsupervised learning allows a network to be
trained without the requirement for pre-classification of the training
set, and free from any biases which might be introduced by a human
trainer. In contrast to most supervised schemes, an unsupervised
network is in principle capable of identifying and grouping previously
unknown or unanticipated object classes within the training set.

Previous applications of Kohonen self-organising maps in the
astronomical arena include image-based star/galaxy classification
\citep{Mahonen95, Miller96} and galaxy morphology classification
\citep{Naim97, Molinari98}, determination of stellar atmospheric
parameters \citep{Fuentes01}, classification of gamma-ray bursts
\citep{Rajaniemi02} and stellar populations
\citep{Hernandez94}. Kohonen maps have also been used to classify
the astronomical literature \citep{Poincot98}.

The self-organising map consists of a set of $M$ {\it neurons} (or
{\it nodes}).
 Associated with each neuron $j$ is a reference vector,
$\bmath{w_j}$. 
When presented with a stimulus in the form of an input
vector, $\bmath{x_i}$, each neuron ``fires'' with a strength that is
related to the similarity between the input and the reference
vector. The reference vector of the most strongly firing neuron in the
map therefore represents the closest match to the input vector.

The neurons are arranged in an $N$-dimensional space, and we have chosen to 
arrange them in a rectangular lattice.  Each dimension
is cyclic, so for example in the case of a 2-dimensional map we
could consider it to have the topology of a torus. Typically $N$ is
chosen to be 1, 2 or 3.
Each neuron $j$ has a Cartesian coordinate $\bmath{r_j}$
in this $N$-dimensional space (hereafter referred to as {\it map space}).

By presenting each member $i$ of some input set to each neuron $j$ in
turn, we can ascertain which neuron most closely matches the input
member and associate the coordinate of the neuron with the input
member. In this way we can assign a location in map space to every
member of the input set.

In our algorithm the reference vectors $\bmath{w_j}$ of the neurons
represent template lightcurves, and the input
vectors $\bmath{x_i}$ the measured lightcurves of the astronomical
objects we wish to classify.

The reference vectors of the neurons are initially assigned random
values. The goal of training the network is to adjust the neuron
reference vectors in such a manner that members of the input set
that are ``similar'' are placed more closely together in map space
than members that are dissimilar.

The network is trained in an iterative fashion by exposing it to a
training set. Each member of the training set is assigned a coordinate
in map space corresponding to the neuron in the map that most closely
matches the member. The reference vectors in the best-matching neuron,
and its close neighbours, are updated by blending them with the input
data. The algorithm continues until all members of the training set
have been classified and have updated the network reference
vectors. In the second and subsequent iterations the network is
repeatedly exposed to the same training set, and the algorithm
proceeds in this manner until some completion criterion 
is met.

The blending process is a key element of the algorithm, as it is this
that allows the network to ``learn'', by adapting the contents of the
neuron reference vectors to match the incoming data. The learning
process is competitive; features in the reference vectors which lead
to good matches with input set members tend to be reinforced. This
reinforcement effect is stronger if the feature is shared by many
members of the training set. As the network learns the coordinates of
the members of the training set in map space evolve, and training
set members which resemble each other tend to self-organise into
clusters within the map space.

\subsection{Data preparation}

Before training of the network commences the training set of
lightcurves are epoch folded on a previously identified period, and
binned to match the cardinality of the the reference vectors in the
network neurons. The binned folded lightcurves are then normalised 
to the brightness range 0 to 1 magnitudes. These folding and 
normalisation steps
yield a training set that is scale-free, and not complicated by issues
of absolute mean brightness, nor the time-scale or amplitude of the
variability present in the astronomical lightcurves. By taking these
steps we ensure that the neural network is classifying on shape alone.

Where a population might be expected to contain members with
lightcurves of similar shape but differing amplitude, one might choose
to normalise on mean brightness alone, in order to allow the network
to distinguish between these classes.

\subsection{Identifying the closest matching neuron}

During the training phase the algorithm must identify which of the
template lightcurves (represented by the reference vectors
$\bmath{w_j}$ in the network neurons) is the closest match to the
lightcurve of the training set member $i$ currently under consideration.

We achieve this by computing for every neuron $j$ the value of a statistic,
$S_{ij, min}$, such that:

\[
S_{ij, min} = \min_{p=1}^{n_w}\left(\sum_q(x_{iq}-w_{jk})^2\right)
\]

where $n_w$ is the cardinality of the input and reference vectors,
$q$ and $p$ allow for all possible phasings between the input and
reference vectors, 
$k=(q+p) \bmod n_w$, and $w_{jk}$ is the $k$th element of the
reference vector of neuron $j$.

The minimisation over $p$ is necessary
in our algorithm as we make no attempt in the data preparation stage
to establish an absolute reference phase for our epoch-folded
lightcurves. We feel that any attempt to do this might prove
unreliable when working with noisy data. Rather we choose to compare
the input lightcurve with the template at all possible phase offsets, and
choose the minimum value of the statistic $S_{ij}$.

The best matching neuron $c$ is that for which $S_{ij, min}$ is
minimised over all of the $M$ neurons in the map.

\subsection{Modifying the network}

Once the best-matching neuron $c$ has been identified its reference
vector and that of its near neighbours are modified as follows:

\[
w_{ij}'=w_{ij}+\alpha(t)h(\bmath{r_c}-\bmath{r_i}, t)(x_i-w_{ij})
\]

where $\alpha$ is the {\it learning rate coefficient}, $h(r, t)$ is the
{\it neighbourhood kernel}, and $t$ is a time coordinate. 
We adopt a Gaussian form for the neighbourhood kernel

\[
h(r, t)=\exp(-r^2/2\sigma^2(t))
\]

The time coordinate $t$ varies linearly with iteration number, from
$t=0$ during the first learning iteration, to $t=1$ during the last
iteration.

\subsection{Evolving $\alpha$ and $\sigma$}

The instantaneous rate at which the network learns is influenced by both
the learning rate coefficient $\alpha$, which controls the degree to
which reference vectors are blended with the input data, and $\sigma$
which determines the effective size of the neighbourhood. In a typical
application of the Kohonen scheme both $\alpha$ and $\sigma$ are
chosen to be monotonically decreasing functions of time, $t$. The
combined effect of such a choice is that the large-scale structure in
the map tends to form early in the training phase, with the finer
details crystallising in the later stages of learning.

In common with Kohonen we have found that the ability of the
network to learn successfully is not very sensitive to the precise
functions chosen to evolve $\alpha$ and $\sigma$. The key requisites
are that $\alpha$ should 
not drop so quickly as to stop the learning process prematurely, 
and that $\sigma_0$ should be chosen such that the
neighbourhood kernel encompasses a sufficient fraction of the map to
allow only large-scale structure to form early in the learning
phase. Subsequently the spatial scale of the learning must be
reduced to allow the network to develop finer structure and settle
into a stable state.
We choose to reduce the width of the neighbourhood kernel linearly
with time. 

We will show that $\alpha$ can evolve linearly or exponentially with
time $t$, or even remain constant throughout the learning phase, with
little impact on the ability of the map to self-organise
in a robust manner.

\section{Performance of the algorithm}
\label{sec-perform}
We have tested our algorithm using two training sets. Firstly we have
generated a sample of 5000 synthetic lightcurves with a distribution
of shapes drawn from four distinct classes. The second training set
comprised 1206 lightcurves from the ROTSE experiment \citep{Akerlof00}.

Our timing tests show that, for a given size of training set, the
algorithm execution time for each iteration scales linearly with the
number of neurons, $M$, and
the cardinality of the neuron reference vectors.

\subsection{Configuration of the network}
The tests presented here were obtained using a two-dimensional network
($N=2$). The two-dimensional network is attractive as it provides
sufficient freedom of expression to allow clusters to form, whilst
ensuring that the results are straightforward to visualise and
interpret (as 2-D images, for example). 

In the traditional Kohonen scheme, as adopted here, the number of
neurons comprising the network is constant through the learning
process and must be determined before training commences. We have
performed tests over a range of network sizes, from $5\times 5$ to
$60\times 60$, and found that the choice does not strongly influence
the network's ability to successfully form clusters. The
key considerations are to ensure that sufficient neurons are available
to allow the network to resolve the subtle differences in the features
of the lightcurves, whilst avoiding a situation where the number of
neurons is comparable to or exceeds the number of members in the
training set.
In post-processing cluster extraction stages (Section~\ref{UmatSect})
it is beneficial to have enough nodes to resolve cluster boundaries
but not so many that the average number of training set members per
neuron drops so low as to make the effect of learning on the map
negligible.
We have found that a network size of $20\times 20$ nodes is suitable
for training sets numbering a few thousand members and is adequate for
real-life datasets such as the ROTSE periodic variables.

The cardinality of the neuron reference vectors, $n_w$, must also be
chosen before training commences, and indeed before the training
lightcurves are epoch folded and binned. We have found that the
performance of the network is not strongly dependent on the choice of
$n_w$, subject to it being large enough to allow the distinguishing
features of the lightcurves to be resolved. We tried values of 32 and 64 
elements for the reference vectors and found no difference in performance 
for the smoothly varying lightcurves in our test dataset. 

The final choices are the ranges and functional forms of 
the $\sigma$ and $\alpha$ learning parameters, and the number of 
iterations carried out over these ranges. In our algorithm the number of 
iterations is not set by testing for a convergence criterion, but rather is 
fixed at the beginning of the run and acts to set the resolution of the 
evolution of $\sigma$ and $\alpha$. 
In all the runs presented in this paper $\sigma$ is initially set to 
half the size of the network and is then reduced linearly to one quarter 
the separation of the neurons (i.e.\   the one-sigma full-width of the 
smoothing Gaussian is reduced from the full size of the network to half the
spacing between neurons). We made a number of different choices for $\alpha$, 
described below.

\subsection{Tests with simulated lightcurves}

\subsubsection{Synthetic dataset}
\label{fakedlcs}
\begin{figure}
\includegraphics[width=84mm]{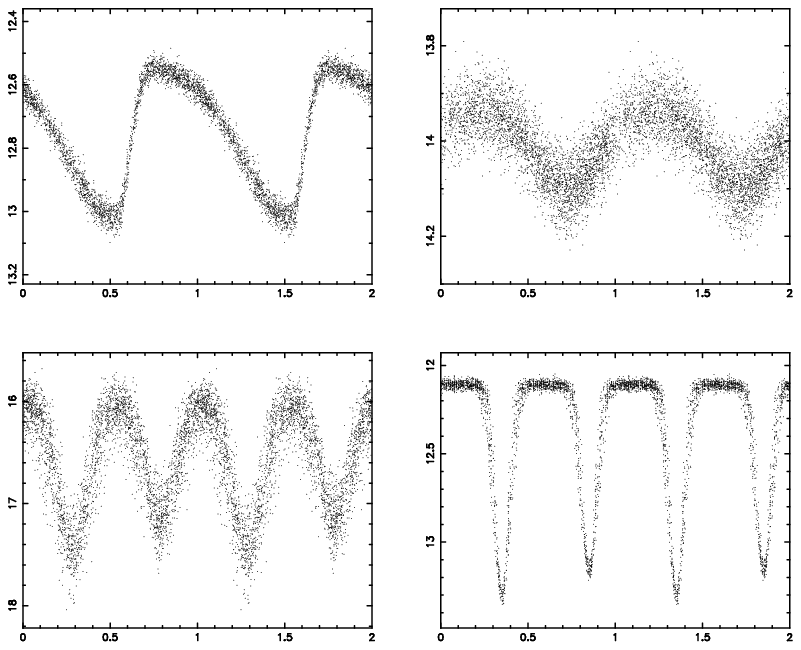}
\caption{Samples of the four classes of simulated lightcurves,
representing type-c RR Lyrae, $\delta$ Scuti and Cepheid types (top
left), type-ab RR Lyrae (top right),
short-period (bottom left) and long-period eclipsing binary (bottom
right)}
\label{fakelcs}
\end{figure}

We simulate lightcurves with the following basic shapes (see
Figure~\ref{fakelcs}):

\begin{enumerate}
\item a single skewed Gaussian peak (representing stars of the RR
Lyrae ab-type family);
\item a single symmetric Gaussian peak (representing stars of the RR
Lyrae c-type, $\delta$ Scuti and Cepheid families);
\item\label{ew} constant brightness with two broad Gaussian dips of equal
width but differing depth (representing eclipsing close
binary stars);
\item as \ref{ew} but with narrower dips (representing well-separated 
eclipsing binaries).
\end{enumerate}

These functions are not intended to represent accurately the real lightcurves 
of such systems, but merely to produce a test data set with the appropriate 
range of behaviour. 

For each synthetic lightcurve the characteristic features of its class
(the width of peaks and dips, and the relative depths of the dips)
were chosen randomly from within ranges chosen to yield profiles
representative of those observed by ROTSE and ASAS 
\citep{Akerlof00,Pojmanski97,Pojmanski02}.

Noise is added to the synthetic lightcurves to represent photometric
measurement errors. The distribution of signal-to-noise ratio in the
set of synthetic lightcurves
is chosen to represent a 
source population with a power-law cumulative distribution in flux 
with a slope of $-1$ (a uniformly distributed source population would have a 
steeper slope of $-3/2$).
In terms of magnitudes our synthetic data have the following mean brightness 
distribution. 

\[
{\rm log} N(<m_v) \propto \frac{2}{5} m_v
\]

\begin{figure}
\includegraphics[height=84mm, angle=90.0]{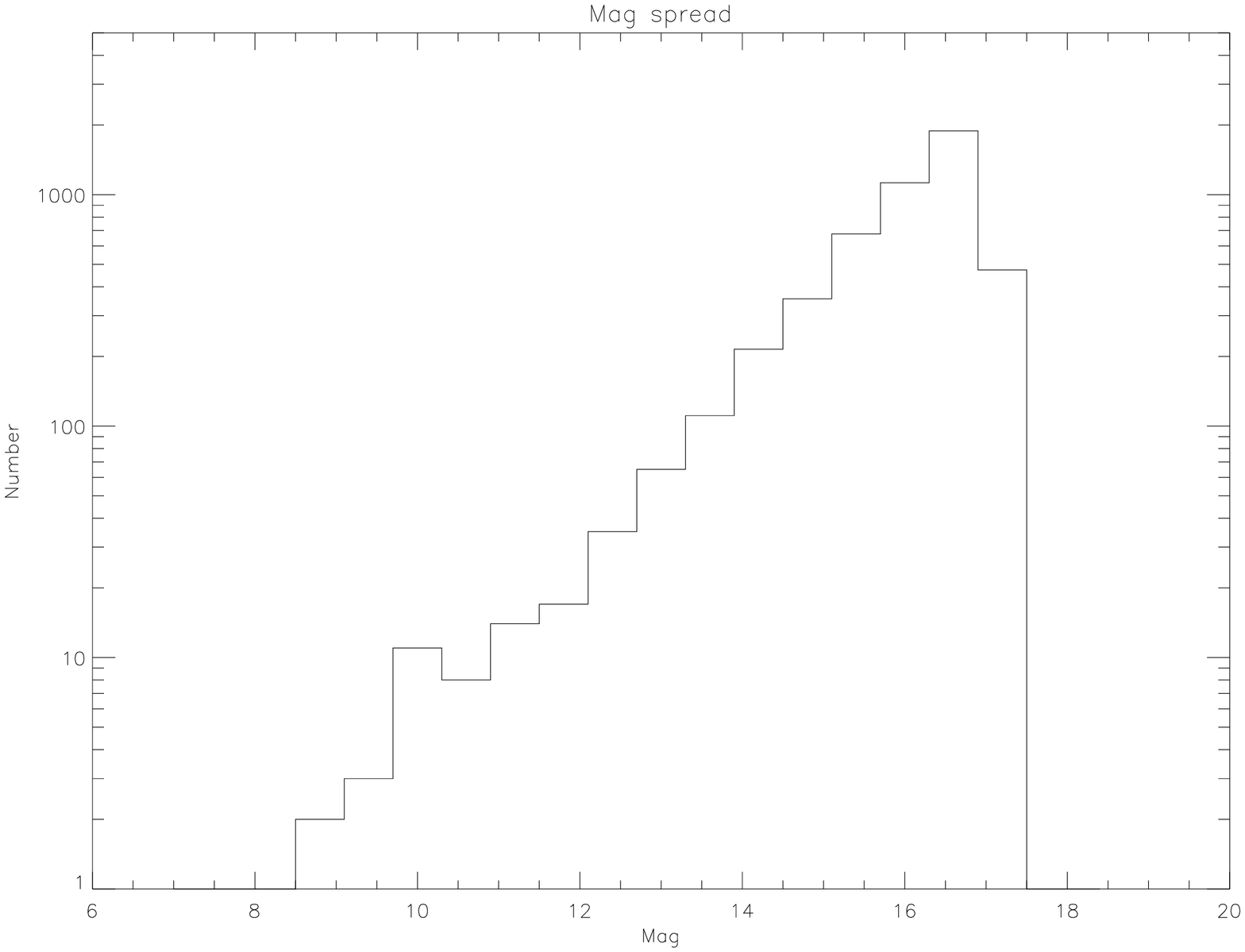}
\caption{The lightcurve brightness distribution for the simulated dataset.}
\label{lcdist}
\end{figure}

Our actual magnitude distribution is presented in Fig.\,\ref{lcdist}.

\begin{figure*}
\includegraphics[width=170mm]{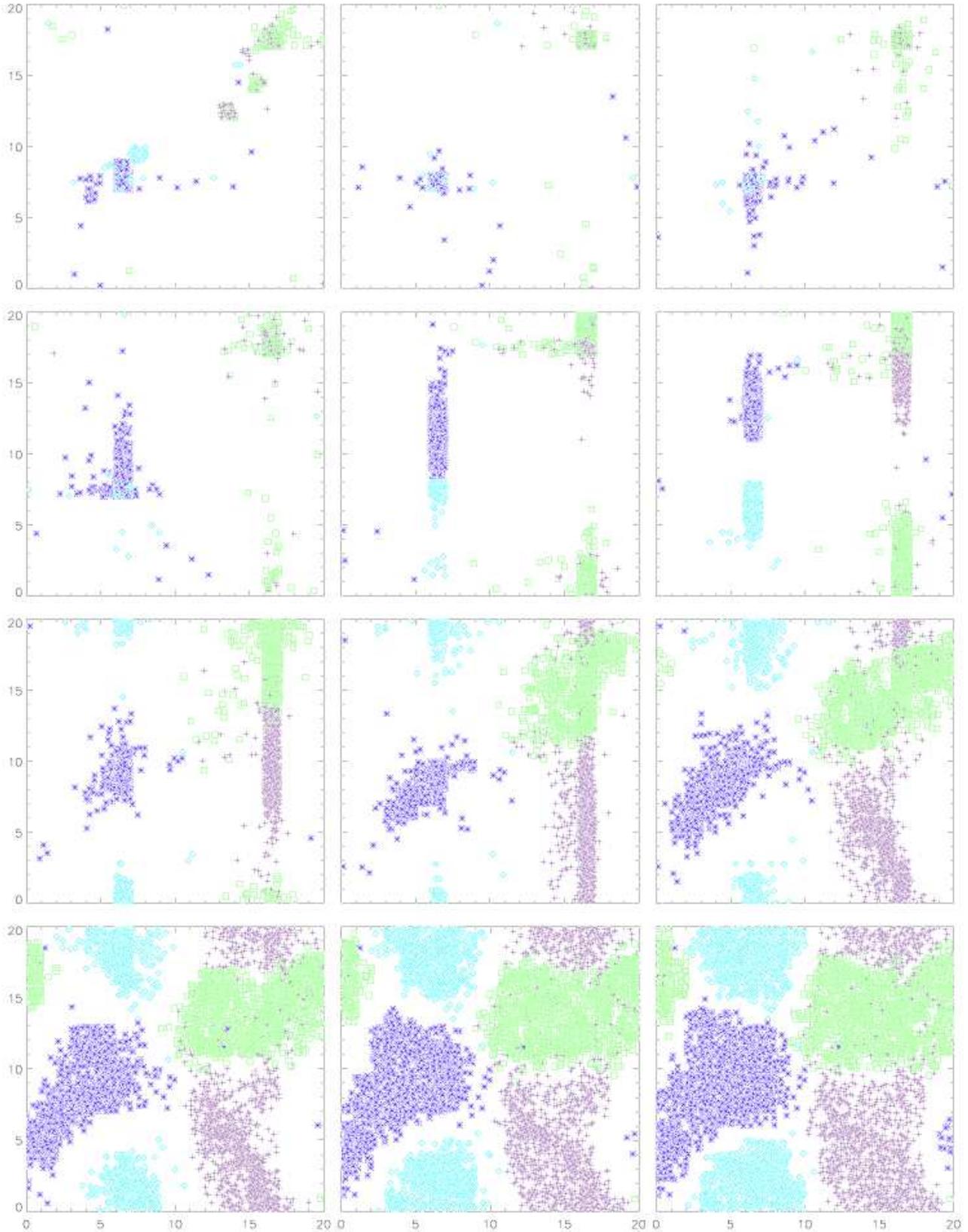}
\caption{Evolution of a network trained with our synthetic data and with 
$\alpha(t)=\alpha_0=0.01$. 
Panels show (from left to right, top to bottom), 
iterations 1, 4, 7, 10, 13, 16, 19, 22, 25, 28, 31, 34. 
The four classes of synthetic lightcurve 
(described in Section~\ref{fakedlcs}) are denoted by (i) squares,
(ii) crosses, (iii) diamonds and (iv) stars. Note that, for the
purposes of visualisation, the positions of all lightcurves have been 
randomised within their host neurons.
}
\label{mapevol}
\end{figure*}

\subsubsection{Learning behaviour of the network}
Figure\,\ref{mapevol} shows the evolution of a typical network. 
In this example the $\alpha$ parameter is held constant at 0.01 throughout. 
The full run had a resolution of 34 iterations. 
Figure\,\ref{mapevol} shows every 
third iteration and, for purposes of visualisation, the positions of all
lightcurves have been randomised within their host neurons.

It can be seen in Fig.\,\ref{mapevol} that the 5000 lightcurves initially
form two tight clusters. 
This is because the synthetic lightcurves naturally divide into two basic 
types (single and double peaked) and because the large smoothing kernel does 
not allow the network to form sufficient structure to 
resolve the lightcurve sub-types. 
The lightcurves are tightly clustered because 
the large smoothing length forces the clusters to strongly repel one another. 
Weak cross-shaped structures are also apparent, and are a natural
consequence of the wrapping of the Gaussian kernel in both dimensions. 
As the network is evolved, and the kernel size is 
decreased, the network becomes increasingly free to develop substructure 
which better represents the the range of shapes of the synthetic lightcurves. 
Between iterations 13 and 16 one of the initial clusters splits into
two (the contact and detached eclipsing binaries), and between 
iterations 13 and 22 the other initial cluster also gradually
splits (the symmetric and skewed Gaussian shapes). By iteration 22 the
map well represents the four types of lightcurve. 
As the kernel size continues to 
decrease the clusters repel each other less strongly and begin to spread to 
occupy more of the available neurons. This is because sharper differences in 
the template lightcurves are now allowed between neighbouring neurons. 
The increasing size of the clusters allow them to better describe the range 
of sub-structure within each type of lightcurve, and smooth variations of 
lightcurve properties are seen across the individual clusters. 
Finally, as the smoothing kernel decreases below the separation of
individual neurons, neighbouring neurons become decoupled and the
clusters can make contact. 
At this point the network has evolved to the point that all of its neurons 
are representing subtly different lightcurve shapes, and learning is complete. 

\subsubsection{Monitoring learning performance}
The learning behaviour of our maps 
motivates us to track three measures which allow
us to quantify the degree of re-organisation which is taking place at
a given time, and the quality of the match between the template
lightcurves and the training set members assigned to them. 

\begin{figure}
\includegraphics[width=84mm]{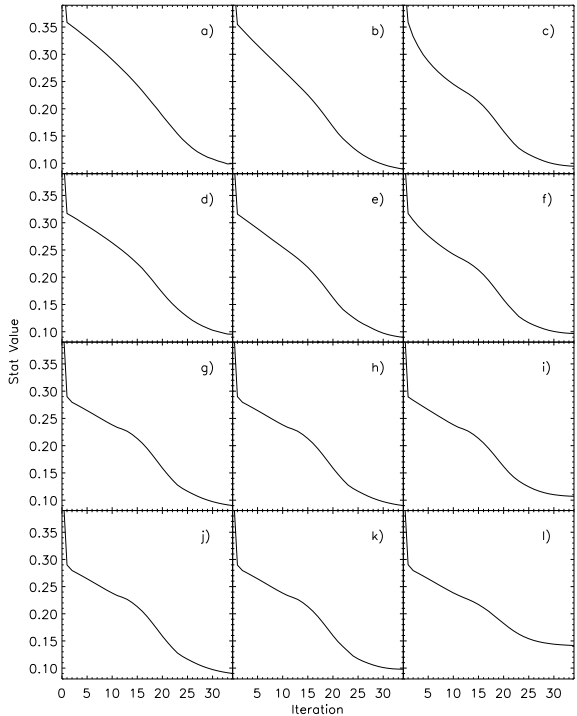}
\caption{The average ``goodness of fit'', \sbar.
In the leftmost
column the learning parameter $\alpha$ is
held constant throughout the learning phase. In the central column
$\alpha$ falls linearly, and in the rightmost column
exponentially. The initial value of $\alpha$ assumes the
value 0.9 (top row), 0.5 (second row), 0.1 (third row) and 0.01
(bottom row).}
\label{sbarstat}
\vspace{1mm}
\end{figure}

\begin{figure}
\includegraphics[width=84mm]{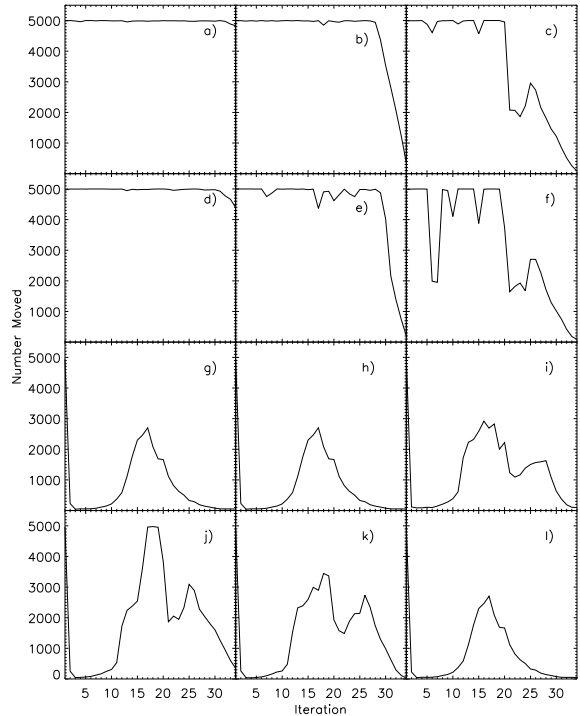}
\caption{The number of training set members that move from one
iteration to the next, \nmove.
Panel order is the same as Figure~\ref{sbarstat}.}
\label{nmovestat}
\vspace{1mm}
\end{figure}

\begin{figure}
\includegraphics[width=84mm]{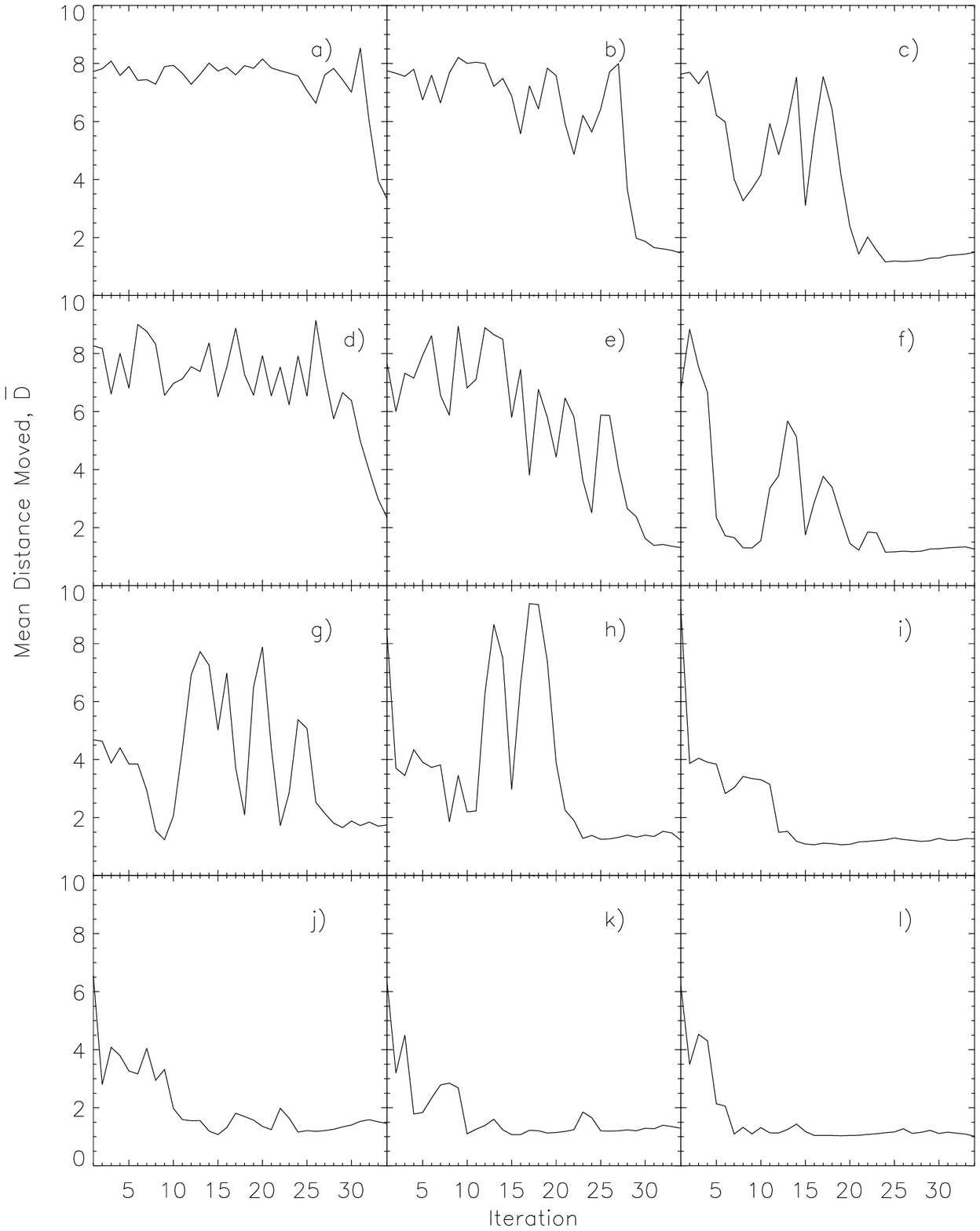}
\caption{The average Euclidean distance
moved by a training set member in map space from one iteration to the
next, \dbar. 
Panel order is the same as Figure~\ref{sbarstat}.}
\label{dbarstat}
\end{figure}

The first of these measures (\sbar) is the value of $S_{ic,min}$ averaged
over all members $i$ of the training set. The statistic $S_{ic,min}$
can be regarded as a measure of the goodness-of-fit between a training
set member and the template lightcurve of the network neuron $c$ which
most closely matches it. By averaging over all members of the training
set we can therefore measure how well the template lightcurves of
the network as a whole have adjusted to represent the input data. We
expect \sbar\ to remain high during early iterations, 
to show a rapid fall once the network becomes free to represent all the 
lightcurve sub-types as separate groups, and then to exhibit a 
slower decline as finer adjustments are made to the template lightcurves.

The second measure (\nmove) is a count
of the number of training set members that have changed location in
map space from one iteration to the next. We expect maximum movement
at times of maximum learning. 

The final measure (\dbar) is the average Euclidean distance
moved by a training set member in map space from one iteration to the
next. Again, we expect maximum movement at times of maximum learning. 

Figures~\ref{sbarstat},~\ref{nmovestat} and~\ref{dbarstat} show the
evolution of \sbar, \nmove\ and \dbar\ as networks are
trained using the synthetic lightcurves described above. 
Several training runs were made, differing only in the
initial value chosen for $\alpha$, and the functional form
with which $\alpha$ evolves. We choose three functional forms for
alpha: a constant value $\alpha(t)=\alpha_0$, a linear form
$\alpha(t)=\alpha_0(1-t)$, and an exponential form
$\alpha(t)=\alpha_0\exp(-\lambda t)$, where the scaling constant
$\lambda$ is chosen such that $\alpha=10^{-3}$ in the final
iteration. 

Taken together these three figures illustrate that the ability of the
map to robustly self-organise is not critically dependent on the
choice of either $\alpha_0$ or the functional form of $\alpha(t)$. In
particular Fig.~\ref{sbarstat} shows that the template lightcurves
adjust to faithfully represent the input data in all cases. All panels
in Fig.~\ref{sbarstat} show a monotonic improvement in the fit
statistic that begins relatively slowly while $\sigma$ is
high, becomes more rapid when $\sigma$ reaches a critical value that
allows all types of lightcurve to be represented by their own
grouping in the network, and finally slows as the maps
approach a steady state at the end of the learning process. 
The final value of \sbar\  is similar in all cases except in 
Fig.~\ref{sbarstat}(l). This run refers to an exponential 
decrease in $\alpha$ from 0.01 to 0.001. It is clear from this panel
that $\alpha$ can drop too fast too soon, thereby ``freezing in''
structure in the network and prematurely ending learning. For this
reason we favour a constant $\alpha(t)$ that allows equal learning at
all stages of the learning process. We also do not use $\alpha$ values
substantially below 0.01
(although lower values may be acceptable as long
as the map is given sufficient time to evolve into equilibrium). 

Figures~\ref{nmovestat} and \ref{dbarstat} reveal more details of the
learning process. Once again, most panels show broadly the same
behaviour, with most learning occurring during the middle of the run, 
and less at the beginning and ends. It is clear, however, that
movement of lightcurves within the map is a function of $\alpha$. 
For $\alpha_0\ltabout 0.2$ the 
evolution of the map appears to be stable, with individual clusters 
moving and evolving slowly and consistently with iteration. 
For larger values of $\alpha_0$, however,
we have noticed that although large-scale structure does indeed form,
that the cluster pattern as a whole continues to migrate rapidly
through map space (Figure~\ref{nmovestat}a, \ref{nmovestat}b, 
\ref{nmovestat}d \& \ref{nmovestat}e). 
In a real-world application this mobility and
constant restructuring would quite likely prove to be undesirable.
For this reason we prefer to choose values of $\alpha_0\ltabout 0.2$.

The rapid learning apparent in the middle iterations in
Figs.\,\ref{sbarstat},\,\ref{nmovestat}\,\&\,\ref{dbarstat} reveals
the presence of a critical value of $\sigma$ which allows the network
sufficient freedom to well represent all classes of lightcurve. This
critical $\sigma$ is likely to take a different value for different
datasets (more classes would require more freedom and a smaller
$\sigma_{\rm crit}$) but nevertheless, in a real-world application,
one might choose to save computing effort by starting $\sigma_0$ at a
value closer to $\sigma_{\rm crit}$. 

There is evidence of some misclassification in the final trained
network in Fig.\,\ref{mapevol}, 
for example between skewed and non-skewed Gaussian
profiles. This is not unexpected as some members of the skewed
Gaussian class will exhibit a very small degree of skewness, and are
difficult to distinguish from the non-skewed class members even when
inspected by eye. Misclassification is difficult to avoid in any
situation where there is a smooth trend in variation of light-curve
shape between two classes, 
emphasising that in real-world applications
shape-based classification would be used alongside other diagnostic
attributes. Nevertheless as a whole the algorithm has performed
remarkably well.

\subsection{A test with real lightcurves}
In order to test the classification abilities of the self-organising
map using real data we chose to train it using lightcurves from the
ROTSE experiment. The training set consisted of 1206 
periodic lightcurves which had been independently pre-classified by other 
means \citep{Akerlof00}. The externally derived classifications were not 
used in training our network, however they did allow us to
assess the ability of the algorithm to group and classify class
members after training is complete.

\begin{figure}
\includegraphics[height=84mm,angle=90.0]{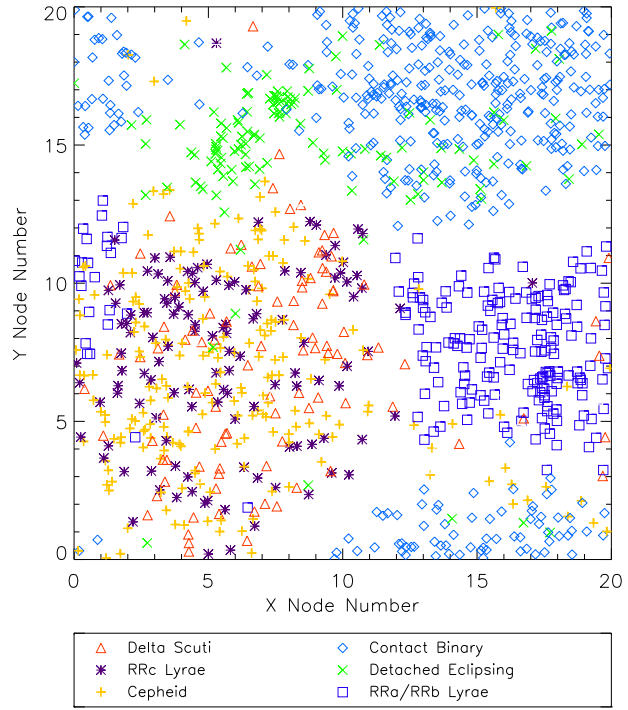}
\caption{The final structure of a network trained using 1206 pre-classified 
ROTSE lightcurves. }
\label{rotseSOM}
\end{figure}

\begin{figure}
\includegraphics[width=84mm,angle=90.0]{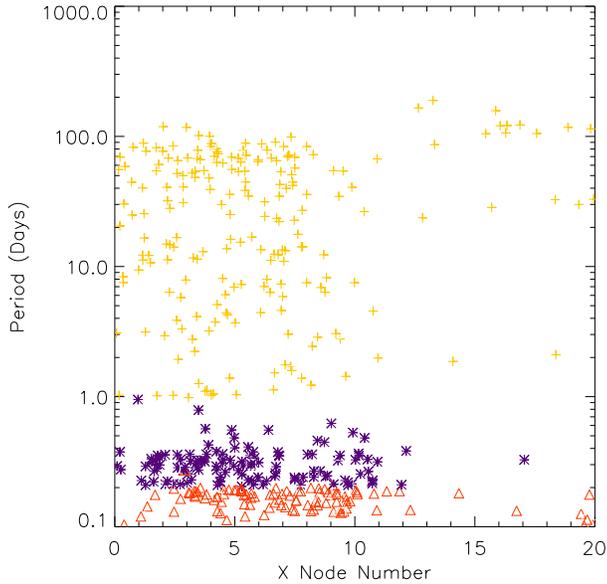}
\caption{The ROTSE map of Fig.~\ref{rotseSOM} collapsed in one dimension and 
expanded with the period of the variable star. Here the $\delta$ Scuti,
Cepheid and type-c RR Lyrae variables, which formed an overlapping group in 
Fig.~\ref{rotseSOM}, have been separated in period. This illustrates how 
additional information can be used to resolve ambiguity in our maps, which 
are based on shape alone. }
\label{rotseSOMPER}
\end{figure}

Figure~\ref{rotseSOM} shows the final state of a network
trained using the ROTSE lightcurves. All parameters were set to the same
values as the run in Fig.\,\ref{mapevol}, in particular $\alpha$ was
held at a value of 0.01 throughout. It can be seen that our algorithm
has successfully differentiated the contact binaries, 
the detached eclipsing binaries, and the type-ab RR Lyrae
systems. 
The remaining three classes in the sample, Cepheids, $\delta$ Scuti and type-c 
RR Lyrae stars,  possess very similar lightcurve profiles 
(indeed the variability is driven by the same underlying physical process), 
therefore the network cannot distinguish between the classes on 
shape alone. 
In such cases, however, the use of additional information can be used to resolve the ambiguity. In Fig.~\ref{rotseSOMPER} we show an example of how this 
additional information can be used. This figure shows the map of 
Fig.~\ref{rotseSOM} collapsed in one dimension and expanded with the 
period of the variable star. The $\delta$ Scuti,
Cepheid and type-c RR Lyrae variables, which formed an overlapping group in 
Fig.~\ref{rotseSOM}, have been separated in period with clearly defined 
cut-off boundaries for each type. 

Some misidentification of lightcurves is apparent in Fig.~\ref{rotseSOM}. 
We have investigated a number of cases and are confident that our maps have 
classified the data correctly. The discrepancies seem to be the result of 
occasional misclassification in the original ROTSE analysis. 

\section{Automated extraction of Clusters}
\label{UmatSect}
Once training is complete and the map has self organised,
it may in some applications be desirable to proceed further to
automatically identify ``clusters'' within the map. Such a cluster
extraction step would then allow the automatic assignment of a
classification to each member of the input set.

\begin{figure}
\includegraphics[width=84mm,angle=90.0]{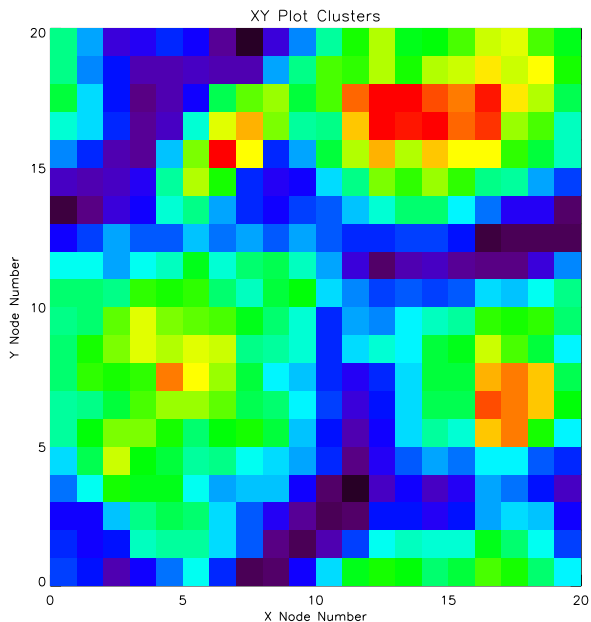}
\caption{A map of the number of training set members assigned to each
  neuron for the network trained with the ROTSE data. The raw map has
  been filtered with a 3x3 top-hat kernel. The colour scale runs from
  black (low occupancy) to red (high occupancy).}
\label{rotseNumDens}
\end{figure}

The most obvious approach to cluster extraction is a direct
examination of the distribution of training set members in map space.
One potential drawback of this approach is
that the Kohonen scheme 
tends
to spread the input members 
fairly uniformly 
throughout the 
map space \citep[Fig.~\ref{mapevol}, see also][]{Kohonen90}. 
This could be a
particular problem in the case of small or noisy datasets, such as the
ROTSE data (Fig.~\ref{rotseSOM}); in such cases the boundaries
between clusters could become difficult to
identify. Figure~\ref{rotseNumDens} shows the number of training set
members assigned to each neuron for the ROTSE data set. A small degree
of smoothing has been applied to this map to reduce noise due to
small-number statistics in the map bins. The clusters and cluster
boundaries are readily visible.

\citet{Ultsch90} outline a different approach to the problem by
calculating what they call the {\it U-matrix}. In essence the U-matrix
attempts to capture the rate of change of shape in the neuron
reference vectors across the map space; within clusters adjacent
neurons will have reference vectors which are rather similar to each
other, whereas at boundaries the differences will be more marked.

The difference between the reference vector of a neuron, $i$, and that
of one of its near neighbours, $j$, can be calculated as:

\[
L = \sum_q{(w_{iq}-w_{jq})^2}
\]

\begin{figure}
\includegraphics[width=84mm, angle=90.0]{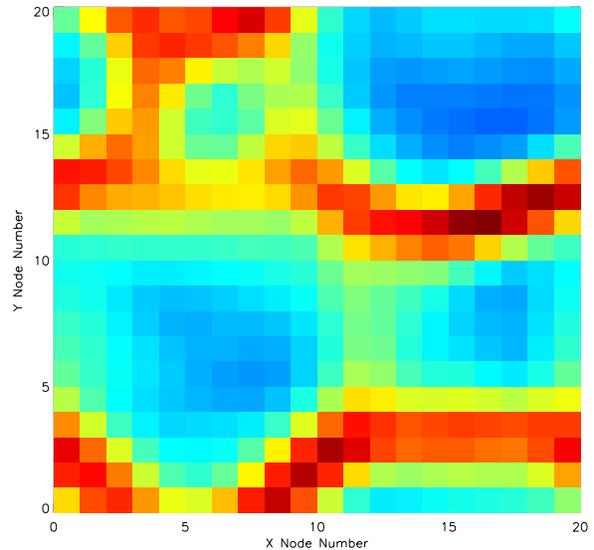}
\caption{The final iteration of the ROTSE data map (shown in
  Fig.~\ref{rotseSOM}) when processed with the U-Matrix
algorithm. The colour scale runs from dark blue (similar reference vectors) to
  red (dissimilar reference vectors).} 
\label{UMatROTSE}
\end{figure}

In a two-dimensional map each neuron will have eight nearest
neighbours, so we calculate $L$ for each neighbour in turn and take
the average. Figure~\ref{UMatROTSE} shows the resulting map calculated
from the same trained network as Figure~\ref{rotseNumDens}. The
U-matrix map also readily reveals the clusters and their boundaries,
and with noticeably less noise than the number density distribution
map.

Both the number density and the U-matrix maps would be suitable for
further processing to identify contiguous areas enclosed within the
cluster boundaries. It is worth noting however that because the
U-matrix is comparing the shape of the templates, rather than number
counts in the training set, the resulting map is less susceptible to
noise in cases where the training set is of limited size.

\section{Conclusions}

We have applied the Kohonen self-organising map scheme to the
classification of singly periodic astronomical lightcurves. Our algorithm has proved
capable of reliably classifying both synthetic and
real lightcurves. The investigation has also shown that the
ability of the map to robustly self-organise is not strongly dependent
on the parameters used to control the learning process. 

We conclude that the self-organising map will prove a valuable addition
to the data-mining toolset for future large time-domain datasets.

\section*{Acknowledgements}
We thank an anonymous referee for helpful suggestions. 
DRB acknowledges the support of the UK Particle Physics and Astronomy
Research Council (PPARC) through the provision of an e-science
studentship. Astrophysics research at the University of Leicester is
also supported through PPARC rolling grants.

\bibliographystyle{/remote/pjw/papers/tex/latex2e/mn2e}
\bibliography{/remote/pjw/papers/tex/bibtex/mn_abbrev2,/remote/pjw/papers/tex/bibtex/refs}

\end{document}